\documentclass{eas}
\usepackage{graphicx}
\runningtitle{AXIAL ROTATION AND TURBULENCE OF RR\,AB STARS}

\begin{document}

\title{AXIAL ROTATION AND TURBULENCE OF RR\,ab STARS: 
THE PETERSON CONUNDRUM REVISITED} 
\author{G. W. Preston}\address{Carnegie Observatories, 813 Santa Barbara Street, Pasadena, CA 91101, USA}
\author{M. Chadid}\address{Universit\'e Nice Sophia-Antipolis, 
Observatoire de la C\^ote d'Azur, UMR 7293, Parc Valrose, 06108 Nice Cedex 02, France}

\begin{abstract}
We calibrate and then use the relation between equivalent width ($EW$) and full--width--half--maximum ($FWHM$) of metallic absorption lines in the spectra of RR Lyrae stars to estimate a new upper limit of 
$V_{rot}sini \leq 6\,km.s^{-1}$ on their axial equatorial rotational velocities, and to derive 
the variations of macroturbulent velocities in their atmospheres during pulsation cycles.  Finally, we present a simple way to estimate macroturbulent/rotational velocity from FWHM of the 
cross--correlation function.
\end{abstract}
\maketitle
\section{Introduction}

The variations of turbulence during the pulsation cycles of classical cepheids 
(Bersier \& Burki \cite{bb}), Breitfellner \& Gillet \cite{bg}), Stifft \& Gillet \cite{sg})) and RR Lyrae stars (Chadid \& Gillet  \cite{cg}), 
Fokin, Gillet, \& Chadid  \cite{fgc}) are well-documented.The definition of turbulence and the procedure for extraction of this quantity 
from stellar line profiles varies from study to study.  However, the disentanglement of turbulence, however defined, from the broadening 
produced by axial rotation is problematic in all of these investigations, which, for the most part, have been concerned with the generation 
of turbulence and its effect on pulsating atmospheres.  This 
problem of disentanglement is well-illustrated in Fig.\,1.  The left panel shows, first of all, that the observed profile of an unsaturated 
line in the spectrum of a typical RR\,ab star is well matched by a Gaussian; there is no obvious rotational signature.  The right panel shows why: 
two convolutions with identical $FWHM$ but very different Maxwell and rotation parameters  
(blue: $\sigma_{M} = 7\,km.s^{-1}$ , $V_{rot} = 0\,km.s^{-1}$; red: $\sigma_{M} = 5\,km.s^{-1}$, $V_{rot} = 6\,km.s^{-1}$) can barely be distinguished in the wings.  
Because rotation, whatever 
its magnitude, is virtually constant during a pulsation cycle, any variation of $FWHM$ with phase must be due to other motions in RR\,ab atmospheres.  
Rotation must be pursued by different means.

Peterson, Carney, \& Latham (\cite{pcl}) found a common minimum value of $FWHM$ near phase = 0.35 derived from cross correlation of metal lines for radial velocity during 
the pulsation cycles of twenty seven RR\,ab stars.  They attributed the variation of $FWHM$ during pulsation cycles to variable turbulence and the minimum to an 
unresolvable combination of turbulence and axial rotation.  Thus, they estimated an upper limit of $V_{rot}sini < 10\,km.s^{-1}$  for the RR Lyrae stars, a result similar 
to that obtained for classical cepheids by Bersier \& Burki (\cite{bb}) .  However, approximately $\sim$\,1/3 of the blue horizontal branch ($BHB$) stars with $T_{eff} < 11500\,K$, 
immediately adjacent to the RR Lyrae domain, have $V_{rot}sini > 14\,km.s^{-1}$  (Behr, \cite{ba} ,\cite{b} , Kinman et al. \cite{ketal}, Peterson, Rood, \& Crocker  \cite{prc} and 
references therein, 
Recio-Blanco et al.  \cite{rb}).  Robust HB theory tells us that all of these stars will traverse the RR Lyrae domain with apparent rotational velocities above the RR Lyrae 
upper limit on their paths to the AGB (Preston \cite{p2}); hence the “Peterson Conundrum”.  We pursue this
puzzle by study of an independent sample of RR\,ab stars chosen to investigate the dynamical properties of RR Lyrae envelopes during their pulsation cycles, 
as described in Preston (\cite{p2}).

\section{ OBSERVATIONS AND PROCEDURES}
\subsection{Observational Data}
The observational data for this study are provided by thousands of echelle spectra of three dozen bright RR\,ab star acquired with the du Pont 2.5\,m telescope of the Las 
Campanas Observatory in the years 2006 -- 2012.  Typical spectral resolution is $R \sim 27000$ at $\lambda5000\, \dot{A}$.  Exposure times never exceed 600\,s ($\sim\,0.01\,P$) and the typical 
signal-to-noise ratio is S/N $\sim$\,20.  For stable RR\,ab stars studied here S/N at nearly all phases can be improved by factors of two or more by co-addition of spectra in
small intervals of phase.  Reductions that produce wavelength calibrated, extractions of duPont echelle spectra are described in detail by For, Sneden, \& Preston (\cite{fsp}) 
and need not be repeated here.  Analysis of this large collection of RR\,ab spectra is a work in progress.
\subsection{Procedure}
The first goal of our analysis is to derive estimates of line--of--sight velocity dispersion caused by the combined effects of macroturbulence and axial rotation after removal of 
all other measurable sources of line broadening.  These are, in the order in which they will be discussed: (1) pressure and/or collision damping, (2) instrumental broadening, 
(3) microturbulence, and (4) thermal broadening.

Microturbulence, produced by motions on length scales smaller than the photon mean free path, intensifies unsaturated absorption lines, and can be derived from spectrum analysis 
(Gray \cite{g} , Sneden \cite{s}).  We use the microturbulence values of For, Sneden, \& Preston (\cite{fsp}) in the calculations that follow.  Macroturbulence, produced by motions on length 
scales large compared to the photon mean free path, broadens all absorption lines.  Macroturbulent velocities must exceed the microturbulent velocities of Richardson's (\cite{rl}) 
eddies in the Siedentopf model of turbulent convection (Woolley \& Stibbs \cite{ws}, $R\ddot{u}diger$ \cite{r}) to which we subscribe. 

We approximate corrections for the various line broadening processes enumerated above by use of the additive property, $\sigma_{a} = (\sigma_{b}^{2} + \sigma_{c}^{2})^{0.5}$, of the Maxwell velocity distribution, 
itself a variant of the Gaussian error function.  We assume that turbulent velocities on all length scales are isotropic throughout the metallic line-forming regions of an RR Lyrae 
star.  We adopt the observable parameter $FWHM_{m}$ as our measure of line broadening.  To begin, we remove the effect of instrumental broadening, characterized by $FWHM_{i}$, from adopted 
average $FWHM_{m}$ of unsaturated lines to obtain $FWHM_{u}$, according to

\begin{equation}
  FWHM_{u} = (<FWHM_{m}>^{2} - <FWHM_{i}>^{2})^{0.5}        
\end{equation}

in which $<FWHM_{i}>$  is an average value described below. We then convert $FWHM_{u}$ to Maxwell velocity dispersion $\sigma_{u}$ using Maxwell dispersion  
$\sigma_{u} = (2 \sqrt{ln2})^{-1} FWHM_{u}$.  All further corrections are made in units of 
line-of sight velocity dispersion, $\sigma$.

The first correction, removal of instrumental broadening, must be made to the average value  $<FWHM_{m}>$ rather than to individual lines, because some measured widths of 
unsaturated lines may be narrower than $<FWHM_{i}>$.  These narrowest 
lines properly belong in the average, but they cannot be corrected for instrumental broadening by the formalism of eq. (2.1).

$<FWHM_{i}>$ was evaluated as follows: $FWHM$ was measured for some three dozen lines in each of thirty four ThAr spectra chosen at random from observations made in 
the years 2006 -- 2012.  From these measures mean values $FWHM_{i}$ and resolution $R$ were calculated for each spectrum.  We use the average value of all these measurements, 
 $<FWHM_{i}> = 11.25 \pm 0.08\,km.s^{-1}$, in equation (2.1) to calculate $FWHM_{u}$ for each RR Lyr spectrum.
 
 \subsection{Estimation of $FWHM_{u}$}
 We estimate  $FWHM_{u}$ by use of a variant of the empirical procedure of Hosford et al. \cite{hrg}, illustrated in Fig.\,2 by a plot of $FWHM$ versus $EW$ for lines of Fe\,I and 
 Fe\,II in 
 the spectrum of the metal--poor subgiant HD\,140283.  We compiled a list of $\sim$\,200 unblended metal lines (mostly due to Fe\,I) on 
 $\lambda4100-5300\,\dot{A}$  from 
 inspection of solar lines 
 identified by Moore, Minnaert, \& 
Houtgast (\cite{mmh}).  We expect that these lines are unblended in our spectra of RR Lyrae stars and in HD\,140283.  We utilize these data to make two estimates of the full width 
half maximum parameter of unsaturated lines, $FWHM_{u}$ during the pulsation cycles of our RR\,ab stars:

(1) The plateau--Hosford et al. identify lines with $EW < 90\,m\dot{A}$ as their ``plateau'' of lines with constant $FWHM$, free of detectable damping wings .  
We calculate the average $FWHM$ in the more restrictive domain $40 \,m\dot{A} < EW < 80\,m\dot{A}$.  Our upper bound of $80\,m\dot{A}$ upper bound seems to better identify the 
onset of a linear increase of $FWHM$ present in our RR\,Lyr spectra.  Furthermore, we do not use lines with measured $EW$ less than  $40 \,m\dot{A}$ because of a ``personal equation''
effect that we encountered in our measurements of weak lines, i.e., those for which central depth is comparable to continuum noise in our spectra.  Accidental negative noise in both 
wings tends to produce broad indistinct absorption features that GWP frequently regarded as unmeasurable. On the other hand, accidental positive noise in both wings tends to produce 
apparently narrow, easily measurable features.  These effects become more important with decreasing spectral resolution and decreasing S/N, and they are particularly bothersome near 
minimum light, when the lines become very broad.  Decisions about measurability will vary from one person to another, hence “personal equation”.  The systematic effect produced by 
these biases, derived from measurements of hundreds of lines in 6 RR\,ab stars at all pulsation phases, is small but significant ($1.07\,km.s^{-1} \pm 0.65\,km.s^{-1}$); 
it appears to vary with pulsation phase, going through a minimum near phase 0.35.

(2) The damping regression at $EW = 80\,m\dot{A}$ produces a second, entirely independent estimate of $FWHM_{u}$.  This regression is based on stronger, more accurately measured lines than those 
that lie on the plateau, and its numerical value depends on the choice of the upper limit of plateau $EW$.  We illustrate the utility of this estimate in Fig.\,3, which is a plot of 
the average differences $FWHM$ (regression at $EW = 80\,m\dot{A}$) $minus < FWHM(plateau)>$ for observations of six RR\,ab stars binned at intervals of $\sim$\,0.05\,P. The average difference,
$0.51 \pm  0.14\,km.s^{-1}$, is comparable to the measurement errors of both quantities.  We adopt the average of these two estimates as our final value of  $FWHM_{u}$ for each 
measured spectrum. 

 \subsection{Measurements of $EW$ and $FWHM_{m}$ for six stable RR\,ab}
 We have adequate observational data to construct $FWHM$ versus $EW$ relations throughout their pulsation 
 cycles for the six stable RR\,ab stars identified as calibration stars (calib) in the first column of 
 Table\,1.  For each star we sorted observations gathered over several years with respect to phase and 
 combined observations made in small phase intervals ($\sim\,0.05\,P$) to increase S/N.  The average number 
 of phase bins per star is 13.  For each spectrum we measured $EW$ and $FWHM_{m}$ for all usable lines in 
 our list by use of the IRAF/splot package.
 
 The $FWHM_{m}$ versus $EW$ diagrams produced by these measurements provide direct evidence of the 
 variation of macroturbulence with pulsation phase.  We illustrate this variation by the montage of 
 diagrams at eight phases during the pulsation cycle of WY\,Ant in Fig.\,4.  The plateau value of 
 $FWHM_{m}$ is high ($\sim\,23\,km.s^{-1}$ ) immediately after maximum light (phase = 0.065) in the 
 top left panel.  It reaches a minimum  ($\sim\,14\,km.s^{-1}$) in the bottom left panel and increases 
 steadily during the remainder of declining light in the right panels   Note the similarity of the 
 plateau value of $FWHM_{m}$ and the damping regression value at $EW = 80\,m\dot{A}$. 
 
 \subsection{The Behaviors of Macroturbulence and Microturbulence in RR\,ab stars}
 We used $FWHM_{i}$ = $11.3 \pm 0.1\,km.s^{-1}$ to remove instrumental broadening via 
 $FWHM_{u} = (<FWHM_{m}>^{2} - <FWHM_{i}>^{2})^{0.5}$ , and converted $FWHM_{u}$ to velocity dispersion 
 $\sigma_{u} = (2 \sqrt{ln2})^{-1} FWHM_{u}$ for each spectrum of each star.  Finally we removed 
 microturbulent and thermal velocity dispersions by use of smoothed variations of these quantities, 
 constructed from data for stable RR\,ab by For, Sneden \& Preston (\cite{fsp}),  to obtain 
 $V_{macrot} = ( \sigma_{u}^{2}-\sigma_{micro}^{2}-\sigma_{th}^{2})^{0.5}$, which we present in the upper left panel of Fig.\,5.  
 Blue symbols denote six stars with [Fe/H] $<$ -1.0; red symbols denote two stars (HH\,Pup and W\,Crt) 
 for which [Fe/H] $>$ -1.0.  The pulsation amplitudes of the two metal-rich stars are larger than 
 average, so it is not clear at present whether their apparently different behavior in Fig.\,5 is an 
 amplitude effect or a metallicity effect.  We defer discussion of this question until the analysis 
 of our metal--rich sample (see Table\,1) has been completed. 

We use the subscript ``macrot'' of the vertical axis label to remind the reader that $V_{macrot}$
contains the combined effects of macroturbulence and unknown axial rotation.  The concurrent behavior 
of microturbulent velocity, $V_{micro}$ for the stable RR\,ab stars of For, Sneden, \& Preston 
(\cite{fsp}) is shown in the upper right panel of Fig.\,5.  The forms of the two variations are 
similar, differing only in scale.  This is shown in the bottom two panels of Fig.\,5, where mean 
values for data in small intervals of phase ($\sim\,$0.05 P) are plotted with superposed sine waves 
that were adjusted to fit the observations in the phase interval 0.1 $<$ phase  $<$ 0.7.  
Macroturbulence and microturbulence 
 vary together.  Their ratio at all phases of the pulsation cycle satisfies the minimal 
 requirement of Richardson (\cite{rl}), Kolmogorov (\cite{k}), and the Siedentopf model (Woolley \& Stibbs \cite{ws}) 
 that $V_{macrot}$ always exceeds $V_{micro}$ .  We offer the ratio of sine waves for $V_{macrot}$ 
 and $V_{micro}$ plotted in Fig.\,6 as an approximate, idealized description of the way a primitive 
 ``spectrum of turbulence'', as  described by two length scales, varies in RR\,ab atmospheres during 
 their pulsation cycles.
\section{Use of $FWHM$ of the cross correlation function as a proxy for macroturbulence}
The derivation of $<FWHM_{u}>$ is laborious, while the $<FWHM_{cc}>$ of the cross correlation 
function can be calculated in a few seconds.  To take advantage of this simplicity, we extend our 
calculations of $V_{macrot}$ to the larger sample of RR\,ab in Table\,1 by calculating the ratio 
 $FWHM_{u}/FWHM_{cc}$ for all spectra of our calibration stars.  Note that $FWHM_{cc}$ contains width
 contributions from line widths of our template star, CS\,22874-009, (Preston \& Sneden \cite{ps}) and 
 from the instrumental width of the duPont echelle spectrograph, in addition to the widths of lines 
 in the individual RR\,ab spectra, i.e., our ratio $FWHM_{u}/FWHM_{cc}$ is specific to our template 
 star and our spectrograph.

 Inspection of the left panel of Fig.\,7 reveals that the ratio $FWHM_{u}/FWHM_{cc}$ varies smoothly 
 during the RR\,ab pulsation cycle with scatter from star to star that scarcely exceeds measurement
errors.  We use the well--defined minimum value of the ratio, $\cong$\,0.40, that occurs near 
phase = 0.38 as a multiplier by which to estimate $FWHM_{u}$ , hence $V_{macrot}$ by the procedures 
of Sect.\,2, using $FWHM_{cc}$
observed   near phase = 0.38 given in the 4th column of Table\,1.  The right panel of Fig.\,7 is a 
histogram of the $V_{macrot}$ values calculated in this manner. 

Three features of this distribution are worthy of mention:

First, the average values of $<FWHM_{u}>$ used to calculate $V_{macrot}$ for the samples [Fe/H] $<$ -1.0 and 
[Fe/H] $>$ -1.0 in Table\,1, are virtually identical, $26.6 \pm 0.57\,km.s^{-1}$ and $27.2 \pm 0.63\,km.s^{-1}$, 
respectively.  Accordingly, we combined the two abundance groups to make the histogram in the 
right panel of Fig.\,7.

Second, the dispersion in $<FWHM_{u}>$ is very small, $\sim 0.6\,km.s^{-1}$ as noted by Chadid \& Preston (\cite{cp}), and the formal 
dispersion for 
$V_{macrot}$ calculated therefrom is only $\sim 0.2\,km.s^{-1}$.  We adopt 
$V_{rot}sini \leq 6\,km.s^{-1}$ as our new lower 
limit on axial rotation of the RR\,ab stars, taken from the average, $6.2 \pm 0.2\,km.s^{-1}$, of the values 
in column 7 of Table\,1 and displayed in the right panel of Fig.\,7.  We know of no other class of stars
that exhibits such a small range of axial rotation.

Finally, if axial rotation were the principal contributor to $V_{macrot}$, its distribution would have 
to be skewed toward smaller values of $V_{macrot}$ as a consequence of random inclinations of rotation 
axes.  In fact, the distribution is slightly skewed toward larger velocities.  We suggest that the 
average rotation of RR\,ab stars is much lower than our formal limit of $6\,km.s^{-1}$.  Our results are 
similar to those found for the classical cepheids by Bersier \& Burki (1996): no skew toward smaller 
velocities and no evidence for rotation.  The senior author (GWP) cannot resist the temptation to 
indulge in ``I told you so!'' by reference to his forgotten paper of a half--century ago entitled 
``The Effect of Rotation on Pulsation in the Hertzsprung Gap'' (Preston \cite{p1}).

\section{Summary}
We have shown how measurements of the full--width--half--maxima of absorption lines in the spectra of 
RR\,ab stars, combined with literature information about microturbulent and thermal velocities, can 
be used to constrain the axial rotations of RR\,ab stars.  We derive a new upper limit 
$V_{rot}sini \leq 6\,km.s^{-1}$, and we argue that our distribution of derived widths suggests a much
lower limit, virtually no rotation at all.  In view of the much larger (seemingly secure) 
rotations reported for many $BHB$ stars immediately adjacent to the instability strip, we conclude 
that, upon entering the instability strip on their paths to the $AGB$, horizontal branch stars either 
expel the angular momentum from their surface layers or hide it in their interiors.  

If all RR\,c stars share the narrow lines of RR\,ab stars, then angular momentum must be lost during entry into the instability strip on the short 
time--scale required for incipient RR\,c pulsation to reach its limit 
cycle $\sim\,100\,P \gg 0.1\,y$ (Stellingwerf \cite{s}).  However, if angular momentum loss occurs more gradually during RR\,c evolution, the time 
scale could be many orders of magnitude longer $\sim\,10^{7}\,y$.  Constraints on RR\,c rotation by measurements of RR\,c line widths in the manner 
reported in this paper will be required to distinguish between these 
possibilities.

In conclusion, we are happy to report that the Peterson Conundrum is still alive and well.  Just like 
the honoree of this symposium, Sylvie Vauclair!

\begin{table*}
 \begin{tabular}{ccccccc}
 \hline
 \hline
 {Star ID} &{P (d)} &{[Fe/H]}& {$FWHM_{cc}$} & {$FWHM_{u}$}&{Vu}&{$V_{macrot}$}\\
 \hline
 & & $[Fe/H]<-1.0$& & &   &\\ \hline
 WY Ant (calib)&0.574&-1.66&26.6&10.7&6.4&6.2\\
BS Aps&0.582&-1.33&26.7&10.8&6.5&6.2\\
XZ Aps (calib)&0.587&-1.57&26.0&10.5&6.3&6.1\\
DN Aqr&0.634&-1.63&27.4&11.1&6.6&6.4\\
SW Aqr&0.459&-1.24&25.9&10.5&6.3&6.0\\
RR Cet&0.553&-1.52&27.2&11.0&6.6&6.4\\
RV Cet &0.623&-1.32&26.7&10.8&6.5&6.2\\
SX For&0.605&-1.62&26.1&10.5&6.3&6.1\\
DT Hya (calib) &0.568&-1.22&26.8&10.8&6.5&6.3\\
V Ind&0.480&-1.50&26.1&10.5&6.3&6.1\\
SS Leo&0.626&-1.83&25.8&10.4&6.3&6.0\\
ST Leo&0.478&-1.29&25.9&10.5&6.3&6.0\\
Z Mic   (calib)&0.587&-1.28&28.1&11.3&6.8&6.6\\
RV Oct (calib)&0.571&-1.34&26.6&10.7&6.4&6.2\\
UV Oct&0.543&-1.61&25.9&10.5&6.3&6.0\\
AT Ser&0.747&-2.05&26.3&10.6&6.4&6.1\\
VY Ser (calib)&0.714&-1.82&27.0&10.9&6.5&6.3\\
V1645 Sgr&0.553&-1.74&26.5&10.7&6.4&6.2\\
W Tuc&0.642&-1.64&27.1&10.9&6.6&6.3\\
CD Vel&0.574&-1.58&26.5&10.7&6.4&6.2\\
AS Vir&0.553&-1.49&26.5&10.7&6.4&6.2\\
\hline
 & & $[Fe/H]>-1.0$& & &   &\\ \hline
 W Crt&0.412&-0.50&26.9&10.9&6.5&6.3\\
DX Del&0.473&-0.56&28.2&11.4&6.8&6.6\\
V445 Oph&0.397&-0.23&28.0&11.3&6.8&6.6\\
AV Peg&0.390&-0.14&26.6&10.7&6.4&6.2\\
HH Pup&0.391&-0.69&26.3&10.6&6.4&6.1\\
AN Ser&0.522&-0.04&27.4&11.1&6.6&6.4\\
ST Vir&0.411&-0.88&27.2&11.0&6.6&6.4\\
UU Vir&0.476&-0.82&26.7&10.8&6.5&6.2\\
\hline
\hline
 
 \end{tabular}
\caption{RR\,ab stars with measured $<FWHM_{cc}>$ and $<FWHM_{u}>$
      near phase 0.38 used to calculate $V_{macrot}$  
\label{tbl-1}}

\end{table*}

\begin{figure}
\begin{center}
      \includegraphics[angle=00,width=1\linewidth]{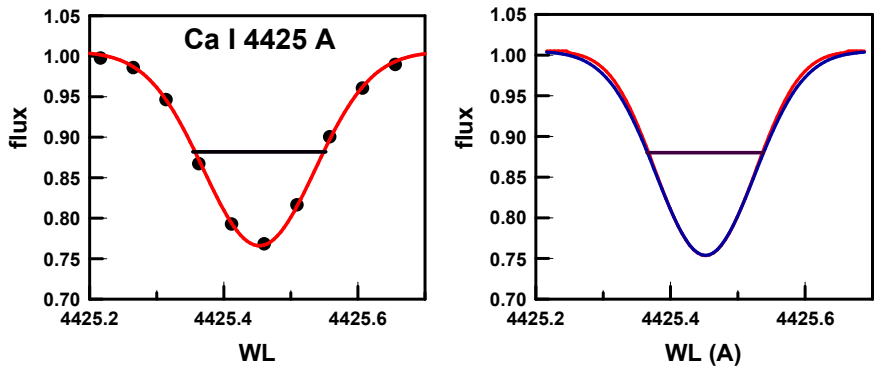}

   \end{center} 
\caption{ Left Panel: Observed profile of $Ca\,I\,4425 \dot{A}$, an unsaturated line in the spectrum of RR\,ab 
WY\,Ant. 
Black circles denote fluxes at the pixel centers of a duPont echelle spectrum.  The profile is 
well--represented by a Gaussian function (red curve).  Right panel: Two convolutions with identical $FWHM$; 
red, Maxwell dispersion = $5\,km.s^{-1}$ , $V_{rot} = 6\,km.s^{-1}$ ; blue, Maxwell dispersion = 
$7\,km.s^{-1}$ , $V_{rot} = 0\,km.s^{-1}$.}
\label{bulge}
  \end{figure}

  \begin{figure}
\begin{center}
      \includegraphics[angle=00,width=1\linewidth]{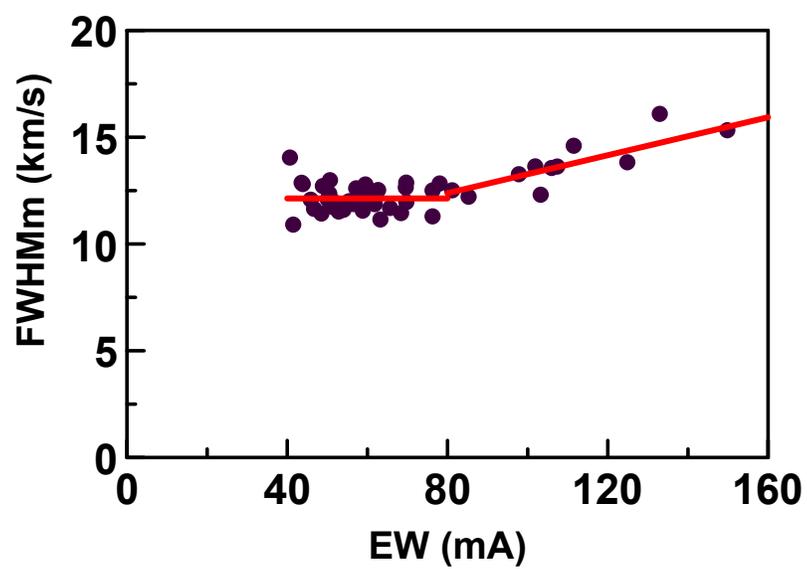}
   \end{center} 
\caption{$FWHM_{m}$  versus $EW$ for lines measured in the duPont echelle spectrum of 
HD\,140283}
\label{bulge}
  \end{figure}
  
  \begin{figure}
\begin{center}
      \includegraphics[angle=00,width=1\linewidth]{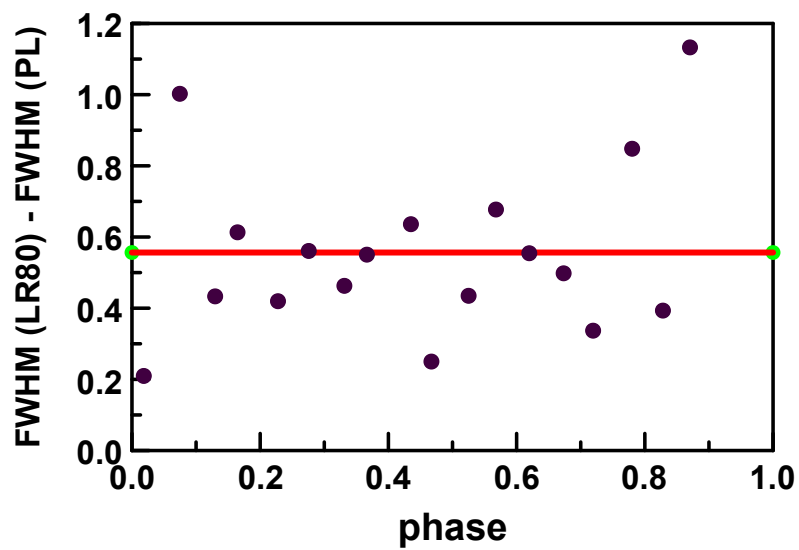}
   \end{center} 
\caption{ The difference $FWHM$ (damping regression at $EW = 80\,m\dot{A}$) minus $FWHM$ (plateau) is 
comparable to the measurement errors of both quantities. }
\label{bulge}
  \end{figure}

 \begin{figure}
\begin{center}
      \includegraphics[angle=00,width=1\linewidth]{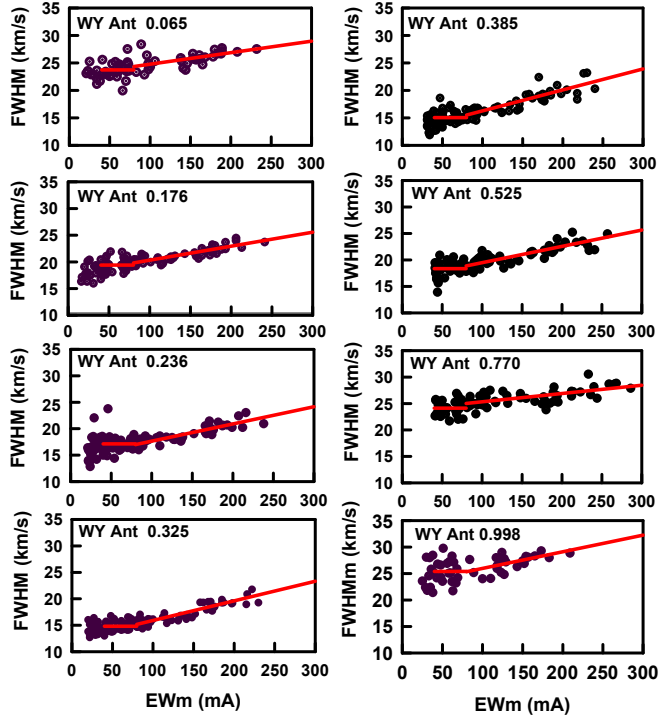}
   \end{center} 
\caption{ Montage showing how plateau  $FWHM_{m}$ ($EW < 80\,m\dot{A}$) varies during the pulsation cycle of  star 
WY\,Ant.  Pulsation phases (0.0 at maximum light) are shown in the upper left corner of each panel.  }
\label{bulge}
  \end{figure}
 
 \begin{figure*}
\begin{center}
      \includegraphics[angle=00,width=1.5\linewidth]{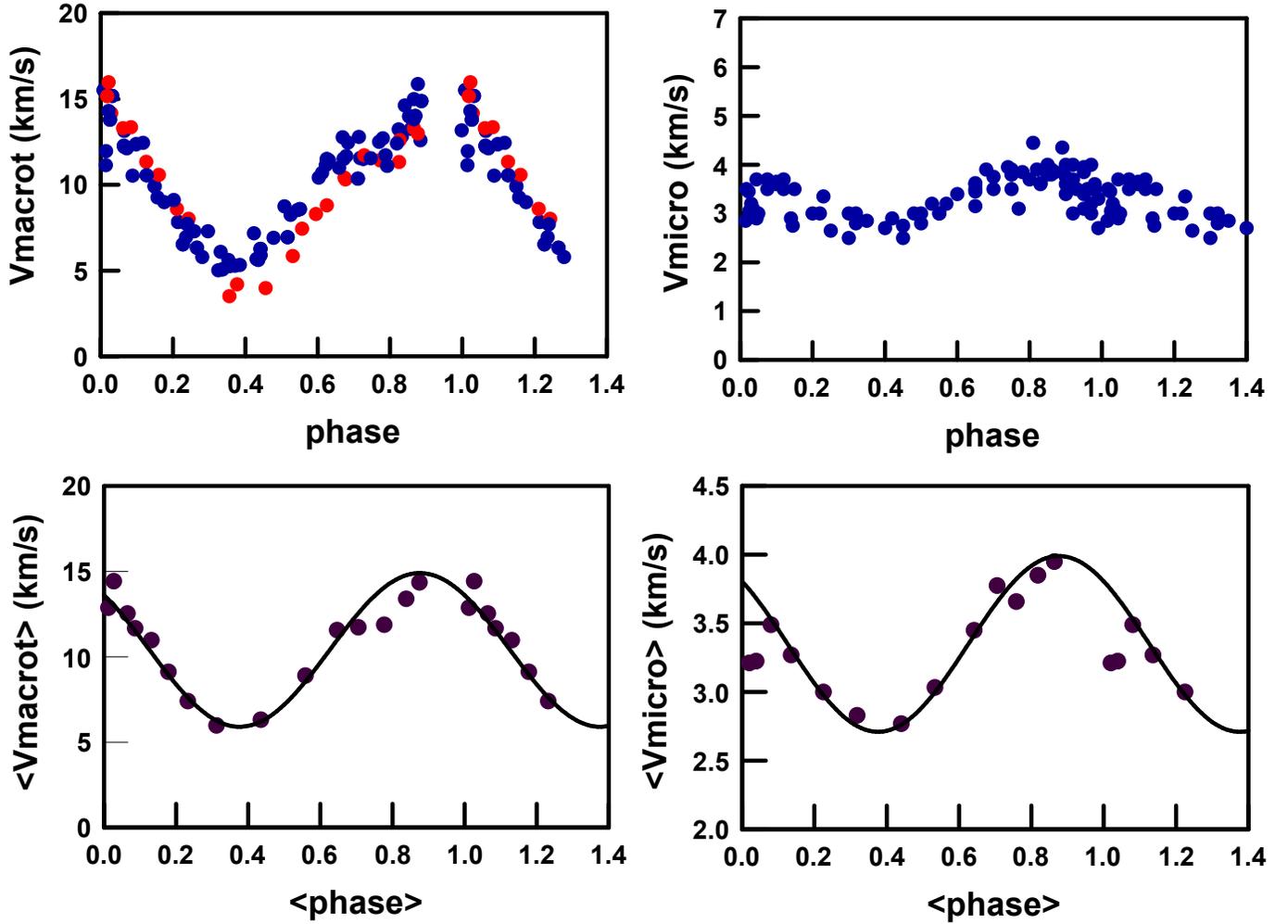}
   \end{center} 
\caption{ Top left: the variation of macroturbulence with phase for six metal--poor (blue) and two metal--rich 
(red) RR\,ab stars.  Top right: the variation of microturbulence in six metal--poor stars.  Bottom panels: 
data for metal--poor in top panels stars are binned into small ($\sim $0.05 P) phase intervals.
Sine waves adjusted to fit data on $0.1 < phase < 0.7 $ are superposed.  }
\label{bulge}
  \end{figure*}
  
 \begin{figure}
\begin{center}
      \includegraphics[angle=00,width=1\linewidth]{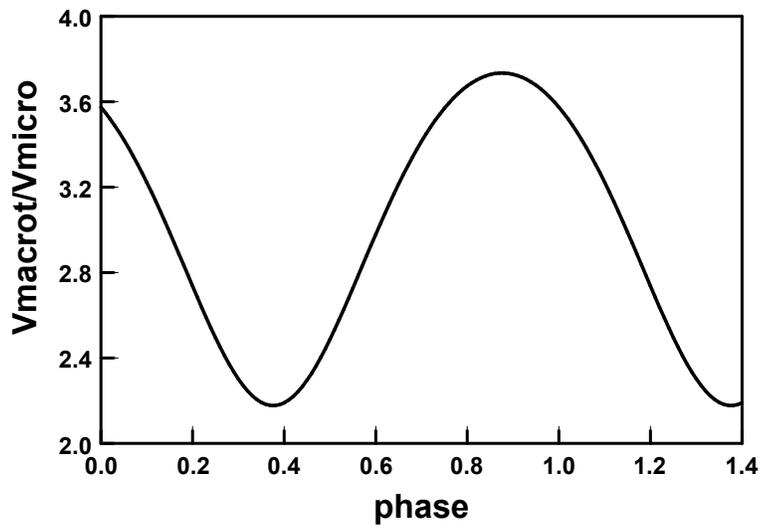}
   \end{center} 
\caption{ The ratio of $V_{macrot}/V_{micro}$ is plotted versus pulsation phase for RR\,ab stars.
  }
\label{bulge}
  \end{figure} 
 
 \begin{figure}
\begin{center}
      \includegraphics[angle=00,width=1\linewidth]{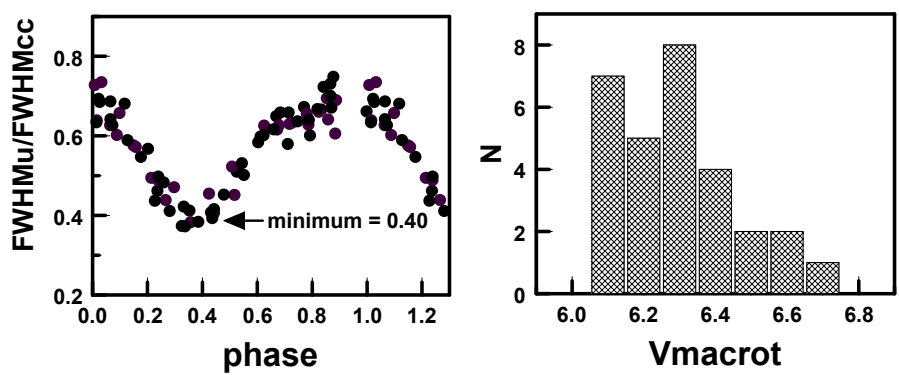}
   \end{center} 
\caption{Left panel: Variation of the ratio $FWHM_{u}/FWHM_{cc}$ with phase defined by the data for the six 
calibration stars in Table\,1.  The variation goes through a minimum near phase 0.38.  Right panel:  
Histogram of $V_{macrot}$ values calculated by use of the minimum value of the ratio.     }
\label{bulge}
  \end{figure}
 
\end{document}